\documentclass[preprint,aps,prb]{revtex4}

\usepackage{graphics}
\begin{document}

\title{Structural study on Pr$_{0.55}$(Ca$_{1-y}$Sr$_y$)$_{0.45}$MnO$_3$ thin films on perovskite (011) substrate}
\author{
Yusuke Wakabayashi$^{1}$\footnote{present address: Graduate School of Engineering Science, Osaka University, 1-3, Machikaneyama, Toyonaka 560-8531, Japan}, 
 Naoko Takubo$^{2}$ \footnote{present address: RIKEN, Hirosawa, Wako 351-0198, Japan},
 Kenjiro Miyano$^{2,3}$,
Hiroshi Sawa$^{4}$
}
\address{
$^1$Photon Factory, Institute of Materials Structure Science, High Energy Accelerator Research Organization, Tsukuba 305-0801, Japan 
\\
$^2$Research Center for Advanced Science and Technology, University of Tokyo, Tokyo 153-8904, Japan 
\\
$^3$CREST, Japan Science and Technology Agency, Honcho 4-1-8, Kawaguchi 332-0012, Japan
\\
$^4$Department of Applied Physics, Nagoya University, Nagoya 464-8603, Japan
}
\date{\today}

\begin{abstract}
Structural study on the photo-switching system Pr$_{0.55}$(Ca$_{1-y}$Sr$_y$)$_{0.45}$MnO$_3$ thin films on perovskite (011) substrate has been made with synchrotron radiation diffraction experiment. The insulating phase of $y=0.20$ sample was found to be an antiferro-orbital ordered phase with large lattice distortion and the ferromagnetic metallic phase of $y=0.40$ film show no distinct lattice distortion from the paramagnetic phase, which are very similar with (Nd,Pr)$_{0.5}$Sr$_{0.5}$MnO$_3$ films. A striking contrast of the orbital ordered phase of $y=0.20$ film to (Nd,Pr)$_{0.5}$Sr$_{0.5}$MnO$_3$ films was its untwinned structure.
\end{abstract}
\maketitle
\section{Introduction}
\label{intro}
Thin films of transition metal oxides have been studied extensively because of their potential to application, and recent report of photo-switching phenomena\cite{Takubo05PRL} is certainly one of the most striking phenomena. Pulsed laser irradiation makes the film metallic, and continuous wave laser irradiation makes it insulating. This phenomenon was found in $y=0.25$ of Pr$_{0.55}$(Ca$_{1-y}$Sr$_y$)$_{0.45}$MnO$_3$ (PCSMO) thin films, a bulk form of which is known as a manganite close to the bicritical point\cite{Tomioka02PRB},  on (011) substrate of the perovskite [(LaAlO$_3$)$_{0.3}$(SrAl$_{0.5}$Ta$_{0.5}$O$_3$)$_{0.7}$] (LSAT). In this study, we report the structural feature of PCSMO thin films on LSAT(011) substrate for $y=$0.20 and 0.40, the lower- and higher- $y$ samples than photo-switching system in order to make clear the electronic states relevant to the photo-switching phenomenon.

\begin{figure}
\resizebox{0.75\columnwidth}{!}{%
  \includegraphics{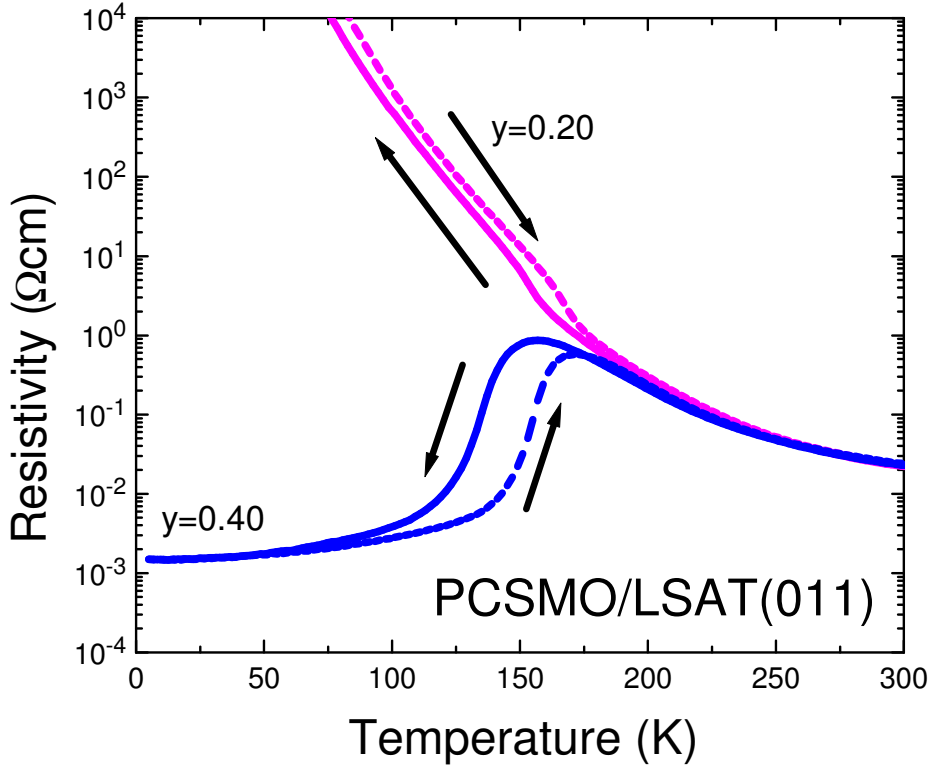} }
\caption{(Color online) Temperature dependence of the electric resistivity.\protect{\cite{Takubo05PRL}}}
\label{fig:rho}       
\end{figure}

The electric resistivity of $y$=0.20 and 0.40 samples\cite{Takubo05PRL} are shown in Fig. \ref{fig:rho}. As shown in this figure, the former is insulator while the latter is metal below 150K. These temperature variations of the resistivity are very similar to those of bulk PCSMO \cite{Tomioka02PRB}. 
According to the bulk phase diagram, bicritical point is at the edge of the $y$ having CE-type charge and orbital ordered insulator phase and the metal phase as their ground states. However, the film system may have different electronic states; For example, the photo-switching phenomenon is observed only in the film system while the photo-induced metallization sustains in a short time in bulk system\cite{Miyano97PRL}. 

The electronic states of manganite thin films are strongly affected by lattice distortion. Epitaxial film technique is widely used for controlling the electronic states through the huge lattice distortion caused by the perovskite (001) substrates \cite{Konishi99JPSJ,Prellier00PRB,Ogimoto01APL,Biswas01PRB,Buzin01APL}, while the strong coupling between the $3d$ orbital degree of freedom and the lattice prohibits to have a first-order metal-insulator transition in such film samples. Recent progress on the study of manganite films is significant: the discovery of the films having first order metal insulator transition\cite{Ogimoto05PRB}. Our previous structural study on (Nd,Pr)$_{0.5}$Sr$_{0.5}$MnO$_3$ films\cite{Wakabayashi06PRL,Wakabayashi08JPSJ} reveals that a part of insulating films on (011) substrate show CE-type state, and some part of them have more homogeneous ferroorbital state \cite{Uozu06PRL,Wakabayashi08JPSJ}. All insulating (Nd,Pr)$_{0.5}$Sr$_{0.5}$MnO$_3$ films have large difference in $b$ and $c$ lattice parameters in their insulating phase, while the two parameters are always the same in the metallic phase. This symmetry breaking makes a twin structure at the metal-insulator transition temperature, and the experimental result also shows all the insulating (Nd,Pr)$_{0.5}$Sr$_{0.5}$MnO$_3$ films have twin.

\section{Experiment}
Epitaxial films were grown using the pulsed laser deposition method\cite{Takubo05PRL,Ogimoto05PRB,Nakamura05APL}
The typical thickness of each of the samples was 80~nm.

 X-ray diffraction experiments were carried out on the BL-4C at the Photon Factory, KEK, Japan. The beamline has a bending magnet, double Si (111) crystal monochromator, and a toroidal mirror as its light source, monochromator, and focusing optics, and is equipped with standard four-circle diffractometer connected to a closed-cycle refrigerator. The photon energy of the x-ray was chosen to 9.5~keV in order to avoid a strong background noise of the luminescence  from Ta in the substrate.

\section{Results and analysis}
\subsection{Insulating film, $y=0.20$}
As shown in Fig. \ref{fig:rho}, the film having $y=0.20$ shows a slight increase of the resistivity at 160-170K, which is attributed to a charge ordering transition\cite{Takubo05PRL}, while the increase of the resistivity is much smaller than other charge ordering manganite films\cite{Ogimoto05PRB,Wakabayashi06PRL}. 
\begin{figure}
\resizebox{1.00\columnwidth}{!}{%
  \includegraphics{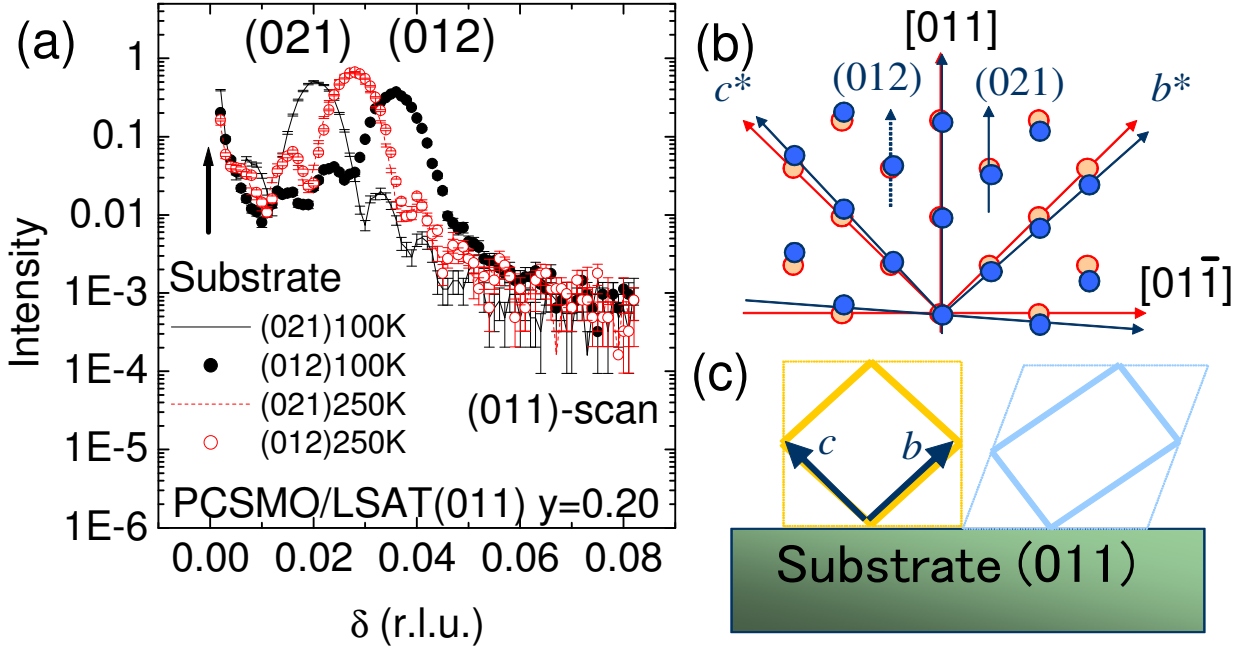} }
\caption{(Color online) (a) Profiles of (012) and (021) Bragg reflections of $y=0.20$ sample measured at 100K and 250K. The peaks shift in the opposite direction by cooling. (b) Schematic view of the reciprocal space. Low temperature phase and high temperature phase are shown by dark (blue) and light gray (red) symbols, respectively. The profiles shown in panel (a) are the intensities measured on the arrows in this panel. (c) Real space image of the film. High temperature phase (left) and the low temperature phase (right). Even in the low-temperature phase, this film has no twin structure.  }
\label{fig:profile}       
\end{figure}
First of all, we measured the change in lattice parameters, which is always observed at either ferro- or antiferro- orbital ordering transition temperatures\cite{Wakabayashi08JPSJ}. Figure \ref{fig:profile} (a) shows the profiles of (012) (plots) and (021) (lines) Bragg peaks measured at 250K and 100K. The Bragg reflections (012) and (021) for the substrate are at $\delta=0$ position. Observed clear fringes indicate the film has homogeneous thickness and good translational symmetry over the irradiated area, i.e., several square millimeters, at below and above the transition temperature. The difference between the two temperatures is apparent. The distance between the substrate peak and the film peak for (021) is much shorter than that for (012) at 100K, while the distances at 250K are almost the same. This difference means the angle of ($01\bar1$) is changed as temperature decreased, as shown in panel (b). The peak profile at 100K shows that the whole film has a single domain structure shown in panel (c), which is unlike to the (Nd,Pr)$_{0.5}$Sr$_{0.5}$MnO$_3$ films on (011) substrates reported in ref.\onlinecite{Wakabayashi06PRL} and \onlinecite{Wakabayashi08JPSJ}. The key factor for this untwinned structure is the as-fabricated tilt of the vector ($01\bar1$) in [011] direction. The angle between this vector and the surface normal, i.e., (011), is 90$^\circ$ for (Nd,Pr)$_{0.5}$Sr$_{0.5}$MnO$_3$ films and 90.03$^\circ$ for $y=0.20$ film at room temperature. In terms of real space, this 0.03$^\circ$ of tilt means a slight amount of deviation from the regular stacking of (011) planes. This deviation breaks the symmetry, and gives the untwinned structure.

\begin{figure}
\resizebox{0.75\columnwidth}{!}{%
  \includegraphics{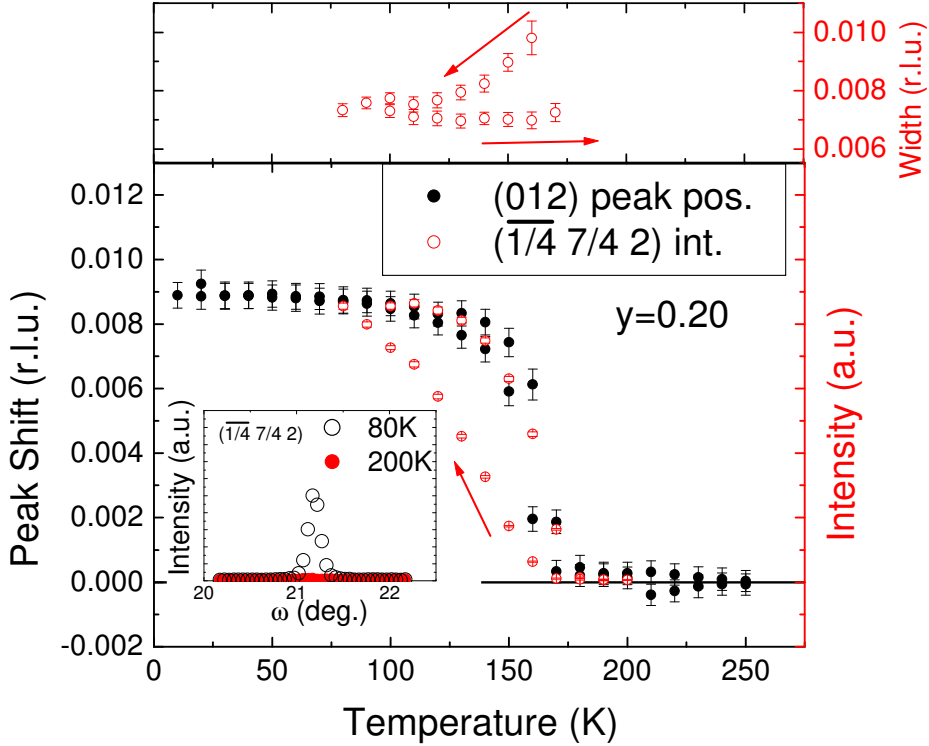} }
\caption{(Color online) (Top panel) Temperature dependence of the width of ($\bar \frac14 \frac74 2$) superlattice reflection in $a$* direction. (bottom panel) Temperature dependence of the amount of the (012) peak shift (closed symbols) and ($\bar \frac14 \frac74 2$) intensity for $y=0.20$ sample. The inset shows the peak profile of ($\bar \frac14 \frac74 2$) superlattice reflection measured at 250~K and 80~K. }
\label{fig:Tdep02}       
\end{figure}

The temperature dependence of the peak shift is shown in Fig. \ref{fig:Tdep02}. 
The peak shift shows a little hysteresis, and the transition temperature is around 160-170K, the same temperature with the slight increase of the resistivity shown in Fig. \ref{fig:rho}. 

Superlattice reflection characterized by the wavevector ($\frac14 \frac14 0$) was observed below 160K as shown in the inset of Fig. \ref{fig:Tdep02}. This reflection is a characteristic of the CE-type charge and orbital ordered phase\cite{Wakabayashi08JPSJ}. The temperature dependence of the superlattice reflection intensity is shown in the figure. Although the significance of the hysteresis of the superlattice intensity is different from that of the Bragg peak shift, the transition temperatures estimated from both superlattice and Bragg reflections are the same. The larger hysteresis observed in the superlattice intensity has larger similarity with the temperature dependence of the resistivity, which shows large hysteresis as shown in Fig.~\ref{fig:rho}. The temperature dependence of the superlattice peak width is also shown in the top panel. The peak width in $a$* direction, or the inverse correration length in $a$ direction, gets larger just below the transition temperature in cooling run, while the peak keeps its sharpness in the heating run. Therefore, the anomaly in resistivity in this film is attributable to the long range ordering in CE-type phase. This result supports the phase diagram given in ref. \onlinecite{Wakabayashi08JPSJ}, small A-site ion radius gives CE-type ordered state. For LSAT substrate, half doped films with A-site ion smaller than 1.355\AA\/ are expected to have charge ordered phase, and those with the ion larger than 1.362\AA\/ are expected to have ferroorbital ordered phase. This sample has the ion of 1.33\AA\/ with the hole concentration of 0.45. The deviation of the hole concentration from 0.5 should make the metallic phase larger, and expected electronic state from the phase diagram for this film is metallic phase or CE-type phase. Therefore, observed electronic state is going along with the phase diagram given in ref. \onlinecite{Wakabayashi08JPSJ}.

\subsection{Metallic film, $y=0.40$}
\begin{figure}
\resizebox{0.75\columnwidth}{!}{%
  \includegraphics{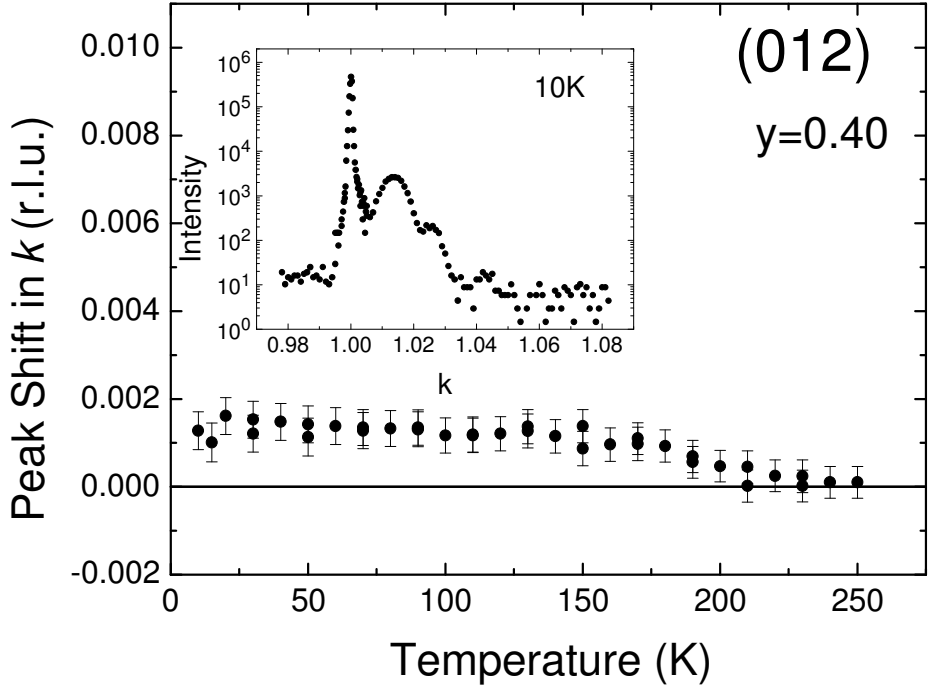} }
\caption{Temperature dependence of the amount of the (012) peak shift for $y=0.40$ sample. The inset shows the peak profile measured at 10K.}
\label{fig:Tdep04}       
\end{figure}

The inset of Fig.\ref{fig:Tdep04} shows the peak profile of (012) Bragg reflection measured at 10~K. It also shows fringes, indicating that the high homogeneity of this film. 
This film has a ferromagnetic metal state as its ground state. For most manganites, the lattice parameters of the paramagnetic phase and those of the ferromagnetic phase are almost the same\cite{Shimomura99JPSJ,Wakabayashi08JPSJ}. As expected, the Bragg peak position of this film shows little change in all temperature region as shown in Fig. \ref{fig:Tdep04}. Although the peak shift is much smaller than $y=0.20$ sample, there is a kink around the transition temperature. This feature means that the metallic phase has smaller thermal expansion coefficient. In addition, no superlattice reflections were observed for this film. Therefore, no charge and orbital ordered insulating phase was found to be coexisted. 

\section{Concluding remark}
Structural study on Pr$_{0.55}$(Ca$_{1-y}$Sr$_y$)$_{0.45}$MnO$_3$ thin films on LSAT (011) substrates was made. The film having $y=0.20$ shows clear CE-type charge and orbital ordered ground state, while $y=0.40$ film shows metallic ground state. Based on the fringes around the Bragg reflections, both films were proved to have homogeneous structure at any temperature. The photo switching film, $y=0.25$, must be really close to the bicritical point as expected from the bulk study.


\end{document}